\documentclass{aa}
\usepackage{txfonts}
\usepackage{natbib}
\usepackage{graphicx}
\usepackage{tablefootnote}
\usepackage{float}
\usepackage{xcolor}
\usepackage{amssymb}
\usepackage{gensymb}

\def\xmm{{\em XMM--Newton}}
\def\sw{{\em Swift}}
\def\inte{{\rm INTEGRAL}}
\def\int{{\rm INTEGRAL}}

\def\approxgt{\mathrel{\hbox{\rlap{\lower.55ex \hbox {$\sim$}}
        \kern-.3em \raise.4ex \hbox{$>$}}}}
\def\approxlt{\mathrel{\hbox{\rlap{\lower.55ex \hbox {$\sim$}}
        \kern-.3em \raise.4ex \hbox{$<$}}}}

\def\ltsima{$\; \buildrel < \over \sim \;$}
\def\lsim{\lower.5ex\hbox{\ltsima}}
\def\gtsima{$\; \buildrel > \over \sim \;$}
\def\gsim{\lower.5ex\hbox{\gtsima}}

\def\hcm {\hbox {\ifmmode $ atom cm$^{-2}\else atom cm$^{-2}$\fi}}
\def\arcmin {\hbox{$^\prime$}}
\def\arcsec {\hbox{$^{\prime\prime}$}}

\def \apjl {ApJL}
\def \apjs {ApJS}
\def \aap {A\&A}

\newcommand{\be}{\begin{equation}}

\newcommand{\ee}{\end{equation}}

\newcommand{\swift}{{\emph{Swift}}}

\begin{document}

\title{Recasting the nature of INTEGRAL hard X-ray transients previously classified as active galactic nuclei}

\author{V.~Sguera\inst{1},   L.~Sidoli\inst{2}  
}
\institute{
$^1$ INAF$-$OAS, Osservatorio di Astrofisica e Scienza dello Spazio, Area della Ricerca del CNR, via Gobetti 101, I-1-40129 Bologna, Italy \\
$^2$ INAF$-$IASF, Istituto di Astrofisica Spaziale e Fisica Cosmica, via A.\ Corti 12, 20133 Milano, Italy \\
}

\offprints{V. Sguera, vito.sguera@inaf.it}

\date{Received  30/07/2025 / Accepted 03/10/2025}

\authorrunning{V. Sguera et al.}

\titlerunning{nature of four INTEGRAL hard X-ray transients}

\abstract{We present new broad-band X-ray results aimed at the identification and characterization of four poorly studied hard X-ray transients discovered by INTEGRAL: IGR J16426+6536, IGR J09446$-$2636, IGR J21268$+$6203, and IGR~J02447+7046. The key properties and X-ray behavior of these sources have remained largely unknown until now. We investigated the temporal, spectral, and energetic characteristics of their hard X-ray outbursts detected above 20 keV by INTEGRAL. In addition, we performed a systematic analysis of unpublished archival soft X-ray observations below 10 keV, enabling a full exploration of their large INTEGRAL error circles in search of the most likely soft X-ray counterparts. Within their arcsecond-sized X-ray error circles, we identified single optical/near-infrared  counterparts for each source. We analyzed their photometric properties to constrain the nature of the systems. Our results show that the X-ray properties of these four transients are inconsistent with the previously proposed extragalactic AGN origin, and instead support a Galactic nature for all of them. Specifically, we propose a very faint X-ray transient  classification for IGR~J16426$+$6536, a nearby flaring star for IGR~J09446$-$2636  and IGR~J21268$+$6203, finally a $\gamma$-ray binary nature for IGR~J02447$+$7046. 
}

\maketitle


 \section{Introduction}

\begin{table*}
\begin{center}
\caption {Characteristics of  the INTEGRAL sources. Coordinates, error radius and significance detection are taken from Bird et al. (2016).} 
\label{tab:igr_info} 
\begin{tabular}{llccccc}
\hline
\hline    
Source       &  Candidate X-ray counterpart   &    R.A. (J2000)  &  Dec (J2000)  &  error radius   &  sig det  & outburst duration    \\
             &                                &                  &              & (90\%, arcmin)  &            &   (days) \\  
\hline    
IGR J16426+6536  &                          &    250.767        &  65.548        &  5.14         &   4.8$\sigma$ & 1-150 \\
                &  XMMLS1 J164303.7+653253  (src. n.1)   &      &                &               &               & \\
                &  4XMM J164246.9+653553 (src. n.2) &   250.695  &  65.598  &              &               &\\
\hline                  
IGR J09446-2636   &                           &  146.152     & -26.565             &  5.45        &  4.5$\sigma$   &  0.6 \\  
                  &   RXS J094436.5-263353  (src. n.1)   &   &                     &              &                 &\\
            &  4XMM J094439.8-263919 (src. n.2)  &  146.166 &   -26.655      &              &                 &\\
\hline                  
IGR J21268+6203  &                            &  321.692              &  62.062   &  4.25         &   5.9$\sigma$  &  3  \\
                  &  2SXPS J212717.7+620603   &  321.823             & 62.101     &               &                &     \\  
\hline                 
IGR J02447+7046 &                                       &     41.178   &  70.775   &  4.78        &  5.2$\sigma$  & 5-30 \\
            &  2SXPS J024443.5+704946           &   41.180            &   70.829 &                &                &   \\
\hline
\hline  
\end{tabular}
\end{center}
\end{table*}

Since its launch in 2002, the \int\ observatory \citep{winkler2003} has played a key role in exploring the hard X-ray transient sky above 20 keV during its long mission, ended in February 2025. Over the past two decades, several INTEGRAL/IBIS catalogs have been published, listing hundreds of hard X-ray transients in the energy range 20--100 keV 
\citep[e.g.][]{Bird2007,Bird2010,Bird2016,krivonos2022,krivonos2018}.
Most transients are Galactic because of INTEGRAL's extensive coverage of the Galactic plane.

\cite{,Bird2010,Bird2016} introduced the bursticity method, which optimizes detection timescales by scanning INTEGRAL/IBIS light curves with variable-sized time windows to maximize source significance. Then, the duration and time interval over which the source significance is maximized are recorded. This technique has proven highly effective, enabling the recovery of dozens of transient sources that would have otherwise been missed. The transients listed in Bird et al. (2016) form a heterogeneous sample with durations ranging from hours to months. However, most remain poorly studied, as the catalogs provide only basic information (e.g., position, average flux, error circle radius), making them ideal candidates for follow-up multiwavelength studies.

To deepen our understanding of these underexplored transients, we are conducting a systematic study using INTEGRAL data (E$>$20 keV) and soft X-ray data (E$<$10 keV). Among the transients under investigation, our attention has been caught by four hard X-ray transients  classified as Active Galactic Nuclei (AGN) in the literature: IGR~J16426$+$6536, IGR~J09446$-$2636, IGR~J21268$+$6203 and IGR~J02447$+$7046. Their temporal and spectral properties in both hard and soft X-rays are largely unknown. In particular, their AGN classification is not robust since a comprehensive study of their nature is still lacking. In fact, no systematic search has been carried out to identify the most likely soft X-ray counterparts of these sources, an essential step to reduce the large positional uncertainty associated with their hard X-ray detections (error circle radii in the range  4$\arcmin$-5$\arcmin$, see Table 1). This requires observations below 10 keV with the X-ray telescopes carried by the {\textit{Neil Gehrels}} \sw\ satellite (hereafter \sw; Gehrels et al. 2004) or \xmm\ \citep{jansen2001}. Such soft X-ray observations are essential to: i) pinpoint the soft X-ray counterpart; ii) reduce the positional uncertainty to arcsecond accuracy; iii) enable optical/infrared counterpart identification and classification. In the following, we summarize the main properties of these four INTEGRAL hard X-ray sources as  known to date.

IGRJ16426$+$6536 is a hard X-ray transient located off the Galactic plane (b$\sim$38$^{\circ}$), discovered with the bursticity method and listed in the latest IBIS catalog of Bird et al. (2016). \cite{ibarra2008} cross-correlated the first \xmm\ slew survey \citep{saxton2008} with the INTEGRAL large positional uncertainty region ($\sim$5$\arcmin$ radius), finding the source XMMLS1 J164303.7+653253 along with two additional ROSAT sources. However, without a targeted soft X-ray observation of IGRJ16426$+$6536 covering its entire error circle, the exact counterpart remained unconfirmed.  Despite this uncertainty, \cite{parisi2008} proposed XMMLS1 J164303.7+653253 as a possible X-ray counterpart mainly because  of the low estimated false detection probability of a random \xmm\ source within the IBIS error circle of IGR J16426+6536 \citep{ibarra2008}. Spectroscopy analysis of the brightest optical USNO source within the  8\arcsec\ (1$\sigma$) \xmm\ error circle  (among the two optical objects present) revealed a narrow-line Seyfert 1 AGN nature with redshift z=0.323 \citep[]{parisi2008,masetti2009}.  \cite{butler2009} came to similar conclusion from their spectroscopic analysis,  although they identified it as a Seyfert 1.5.
Because of this, in the literature IGR~J16426$+$6536 has been assumed to be an AGN, and as such considered for AGN population studies (\cite{panessa2011}; \cite{molina2013}). However, the proposed AGN classification remains highly uncertain.

IGRJ09446$-$2636 is a hard X-ray source firstly reported in the IBIS survey by \cite{krivonos2007}, located off the Galactic plane (b$\sim$20$^{\circ}$). It is also listed in the IBIS catalog of Bird et al. (2016) as a transient source detected with the bursticity method.  Krivonos et al. (2007) and \cite{sazonov2007} cross-correlated the INTEGRAL error circle (5$\arcmin$.4 radius) with the ROSAT catalog and proposed the X-ray source 1RXS J094436.5$-$263353 as counterpart. Based on the optical spectrum from the 6dFGS archive, the ROSAT source was classified as Seyfert 1 AGN with redshift z=0.14 
(\cite{masetti2008}; Krivonos et al. 2007; Sazonov et al. 2007). Clearly, the  proposed AGN nature of IGRJ09446$-$2636 remains tentative.  To date, no soft X-ray investigation of its large INTEGRAL error circle has been performed, hence the true counterpart  is still unconfirmed.

IGR~J21268$+$6203 is a hard X-ray transient discovered with the bursticity method and listed in the IBIS catalog of Bird et al. (2016). Its X-ray properties are unknown,  it has an  error circle radius of  4$\arcmin$.3 and lies slightly off the Galactic plane (b$\sim$8$^{\circ}$). 
\cite{maiorano2011} cross-correlated the INTEGRAL large positional uncertainty  with infrared and radio catalogs. Among the many sources found inside of it, they arbitrarily considered  as candidate counterparts two unidentified sources, one radio and  one infrared, separated by 24\arcsec. Maiorano et al. (2011)  proposed a candidate AGN nature for IGR~J21268$+$6203 based on their radio flux and infrared photometry, respectively.  However, the true X-ray counterpart  is still unconfirmed 
since no soft X-ray investigation of the entire INTEGRAL error circle has been performed to date.

IGR~J02447$+$7046 is a hard X-ray transient discovered with the bursticity method and listed in the latest IBIS catalog of Bird et al. (2016). Its detailed X-ray properties remain unknown. The source is located slightly off the Galactic plane (b$\sim$10$^{\circ}$), with an  error circle radius of 
5$\arcmin$.45 (90\% confidence). \cite{masetti2013}  cross-correlated the INTEGRAL large error circle at 99\% confidence (6$\arcmin$.8 radius) with X-ray and radio catalogs. Among many sources found inside of it, they arbitrarily selected as counterpart a source located on the edge of the large error circle. Optical spectroscopy classified it as a Seyfert 1.2 AGN with  z=0.306. However, this proposed AGN classification   remains tentative since  no soft X-ray investigation of the error circle region of  IGR~J02447$+$7046 has yet been performed to date.


\section{Observations and data reduction}
In the following, we underline the X-ray missions and data used to study the four hard X-ray sources. We checked that no other X-ray data from additional X-ray missions are available

\subsection{INTEGRAL}

Although the four IGR sources are included in the IBIS catalog (Bird et al. 2016), a dedicated re-analysis of the INTEGRAL archival data is required to extend the temporal baseline of the hard X-ray light curves from satellite revolution 1,000 (2010) to revolution 2,000 (2018), and to derive their spectra.

The temporal and spectral behavior of the four hard X-ray transients was investigated above 20 keV using the ISGRI detector \citep{lebrun2003}, the lower energy layer of the IBIS coded-mask telescope \citep{ubertini2003} onboard \inte~ (Winkler et al. 2003), which operates in the energy range 15 keV -1 MeV. \textrm{INTEGRAL} observations are divided into short pointings (Science Windows, ScWs), each typically lasting $\sim$2,000 seconds.

The IBIS/ISGRI public data archive (revolutions 30 to 2,000; January 2003 to September 2018) was systematically searched for hard X-ray activity from the four transients.  IBIS/ISGRI flux maps were generated for each ScW in several energy bands (18–60 keV, 17–30 keV, 20–40 keV) using the latest Offline Scientific Analysis software (OSA 11.2). Count rates at the source positions were extracted from individual ScW maps to build long-term light curves with a time bin of  $\sim$2,000 s. To systematically search for transient hard X-ray activity, we applied the bursticity method (Bird et al. 2010, 2016), which optimizes the detection time-scale by scanning the IBIS/ISGRI light curves with variable-sized time windows to identify the highest source significance, on timescales ranging from 0.5 days to months/years. For each outburst detected, this analysis automatically provides its time interval and energy band of maximum significance. As next step,  we performed a targeted  analysis of each recorded outburst 
by extracting and analysing the spectrum (a systematic uncertainty of 1\% was added to all extracted IBIS/ISGRI spectra) as well as by constraining the outburst duration through investigation of  the source light curve. To this latter aim, the data set includes only ScWs where the sources were within the instrument field of view (FoV) at an off-axis angle $<$12$^{\circ}$, to avoid systematic errors due to poor response modeling at large off-axis angles.

\subsection{\xmm}

The \xmm\ data were reprocessed using version 21 of the Science Analysis Software (SAS), following standard data reduction procedures and using the appropriate calibration files. Response matrix files (RMFs) and ancillary response files (ARFs) were generated using the SAS tasks \texttt{rmfgen} and \texttt{arfgen}, respectively. Background spectra were extracted from source-free regions in close proximity to the target position.
Event pattern selections were applied as follows: patterns 0–4 for EPIC-pn, and 0–12 for MOS cameras.

Spectral fitting of the EPIC (pn, MOS1, MOS2) data was performed simultaneously using \texttt{XSPEC} \citep{Arnaud1996}.  Interstellar absorption was modeled using the \texttt{TBabs} model, with photoelectric cross-sections from \cite{Verner1996} and elemental abundances from \cite{Wilms2000}. Spectral uncertainties were estimated at the 90\% confidence level.
Uncertainties on  X-ray fluxes were estimated using the \texttt{cflux} convolution model in \texttt{XSPEC}.  Cash statistics \citep{Cash1979} (as implemented in \texttt{XSPEC})  was adopted in spectral fitting. Table~\ref{tab:main_outbursts}  reports the summary of the observations analyzed here. 

\begin{table}
\caption {Summary of  \xmm\ observations}
\label{tab:main_outbursts} 
\begin{tabular}{cccc}
\hline
\hline    
source             &     Obs ID      &    start time         &    exp    \\  
                   &                 &    (UTC)              &    (ks)  \\  
\hline    
IGR J16426+6536   &   0670390401     &  2011-05-06 07:06:21  &    22     \\
IGR J09446-2636   &   0761112901     &  2015-06-11 09:20:16  &     18.7  \\       
\hline
\hline  
\end{tabular}
\end{table}


\subsection{The \sw\ observatory}

The observational data performed by the X-Ray Telescope (XRT; \citealt{Burrows2005}) onboard the 
\swift\ Observatory \cite{Gehrels2004} were retrieved from the High Energy Astrophysics Science Archive Research Center (\textsc{HEASARC}) and processed using standard reduction procedures with the \textsc{xrtpipeline} tool. Table 3 reports the summary of the  observations analyzed here.
Cash statistics \citep{Cash1979} (as implemented in \texttt{XSPEC})  was adopted in spectral fitting.
Upper limits were estimated using the tool \textsc{sosta} in \textsc{ximage} on the XRT sky images together with their appropriate exposure maps.
 
\begin{table}
\caption {Summary of the \sw/XRT observations}
\label{tab:main_outbursts_sw} 
\begin{tabular}{cccc}
\hline
\hline    
source             &     Obs ID      &    start time         &    exp    \\  
                   &                 &    (UTC)              &    (ks)   \\  
\hline    
IGR J16426+6536   &   00041799001     &  2010-10-18 11:40:00  &    3.1     \\  
IGR J16426+6536   &   00041799002     &  2010-10-21 07:17:00  &     3.6  \\    
\hline 
IGR J09446-2636   &   00047885001     &  2012-12-18 13:40:00  &     1.1  \\
IGR J09446-2636   &   00041797001     &  2010-10-04 00:37:00  &     0.9  \\
IGR J09446-2636   &   00041797002     &  2010-10-03 13:40:00  &     4.0 \\
\hline 
IGR J21268+6203   &   00045405001     &  2011-04-05 20:53:00  &     0.9  \\ 
IGR J21268+6203   &   00045405002     &  2011-04-08 08:18:00  &     0.8  \\ 
IGR J21268+6203   &   00045405003     &  2011-06-24 17:51:00  &     1.6  \\ 
IGR J21268+6203   &   00045405004     &  2011-07-09 07:46:00  &     0.5  \\ 
IGR J21268+6203   &   00045405005     &  2011-07-16 20:53:00  &     1.8  \\ 
IGR J21268+6203   &   00045405006     &  2011-07-17 20:53:00  &     5.4  \\ 
\hline 
IGR J02447+7046   &   00033613001     &  2015-01-30 05:57:00  &     4.5  \\  
IGR J02447+7046   &   00032222004     &  2012-02-21 01:01:00  &     0.3  \\
IGR J02447+7046   &   00032222003     &  2012-02-19 10:31:00  &     1.7  \\
IGR J02447+7046   &   00032222002     &  2012-02-17 15:11:00  &     0.3  \\
IGR J02447+7046   &   00032222005     &  2016-10-26 03:37:00  &     3.6  \\ 
\hline
\hline  
\end{tabular}
\end{table}


\begin{figure*}
\begin{center}
\includegraphics[height=9.4 cm, angle=0]{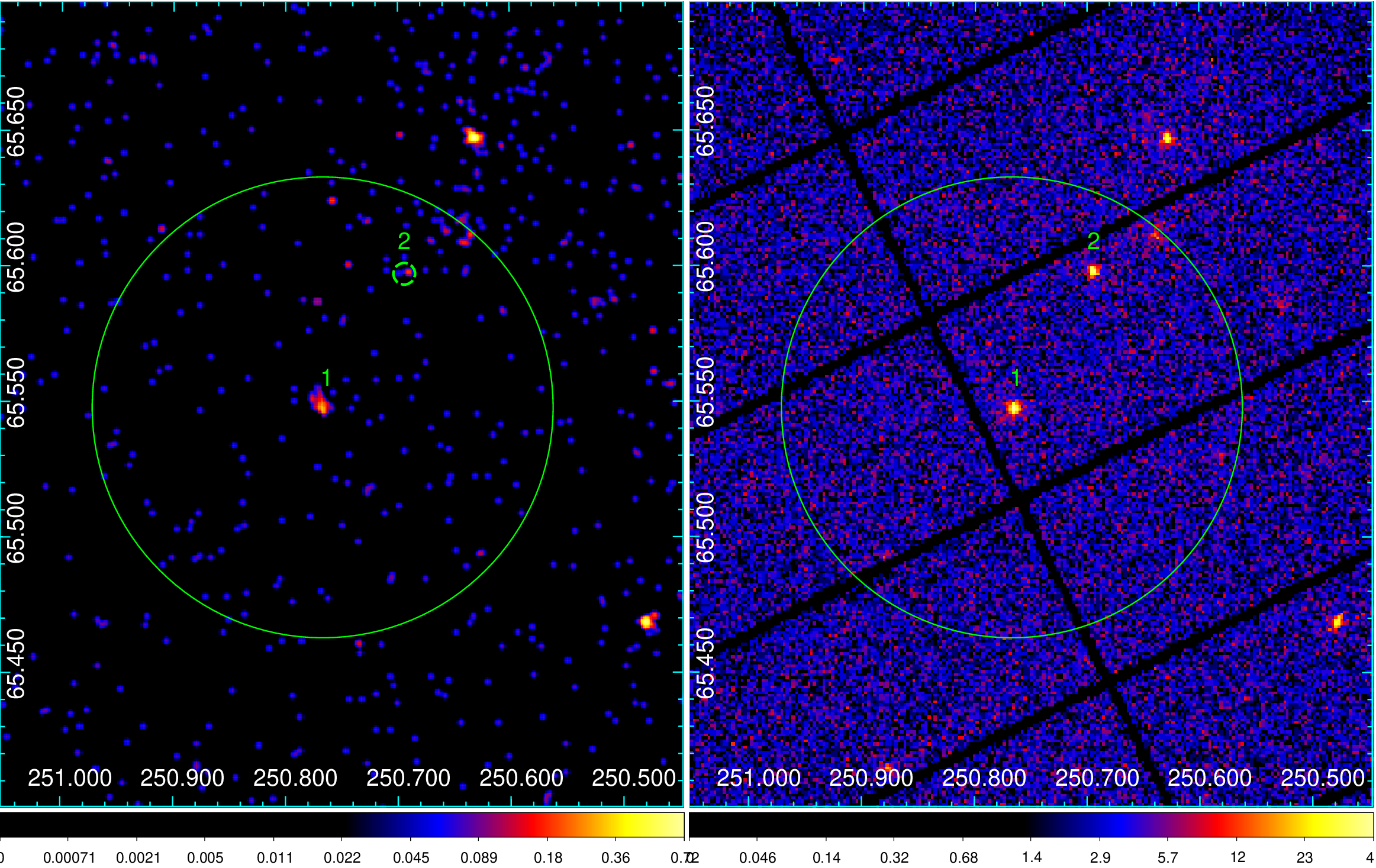}
\caption{\xmm\ EPIC pn (right) and \sw/XRT (left) image  of IGR J16426+6536 sky region 
with superimposed the ISGRI  error circle (90\% confidence). Two X-ray sources (n.1 and n.2) are detected inside of it in the \xmm\ image. Conversely, only X-ray source n.1 is detected in the \sw/XRT image, where the small dashed circle marks the position of the undetected X-ray source n.2.} 
\end{center}
\end{figure*}

\section{Data analysis and results}
\label{sect:data}

\subsection{IGR J16426+6536}

\subsubsection{Hard X-ray observations}

From our analysis of the 15 years IBIS/ISGRI data archive using the bursticity method, we found that IGR J16426+6536 was detected in outburst only once. The hard X-ray activity began on 22 June 2004 at approximately 02:30 UTC (satellite revolution 206).  The source remained within the IBIS/ISGRI FoV until 22 June 2004 at 19:15 UTC (effective exposure of 9.5 ks). It  was best detected in the 20-40 keV energy band, with a significance of $\sim$5.5$\sigma$. The average X-ray flux  was 
8.4$\pm$1.5 mCrab (20-40 keV). Unfortunately, the IBIS/ISGRI temporal coverage of the outburst is limited, as the source moved out of the FoV after 22 June 2004. When it re-entered the FoV about five months later, on 21 November 2004 during revolution 257, it was no longer active, as no significant detection was achieved (20–40 keV). Therefore the duration of the outburst can be constrained in the range 1-150 days, contrary to a previous report in the literature that estimated it as only 0.7 days (Bird et al. 2016). Hence, the INTEGRAL detection likely corresponds to the rising phase of the outburst.

We extracted the average IBIS/ISGRI spectrum of the source during the outburst. The best fit is obtained using 
a power law model ($\chi^{2}_{\nu}$=0.93, 4 d.o.f.)  with photon index $\Gamma$=2.8$\pm$1.5. 
The average 18--60 keV (20--40 keV) flux is 5.5$\times$10$^{-11}$ erg cm$^{-2}$ s$^{-1}$ (3.5$\times$10$^{-11}$ erg cm$^{-2}$ s$^{-1}$).

\subsubsection{Soft X-ray observations}

Figure 1 (right) shows the \xmm/EPIC pn image (0.2-10 keV) of the IGR J16426+6536 sky region, obtained from observation ID 0670390401 (see Table 2). The image is overlaid with the ISGRI error circle. Thanks to this observation, it is now possible to investigate the entire ISGRI error circle in the X-ray band in order to identify the most likely X-ray counterpart. Within it, the X-ray source n.1 corresponds to the AGN previously proposed in the literature as counterpart. Notably, an additional X-ray source (n.2) is also detected, which has never before been considered as potential counterpart.

Figure 1 (left) displays the \sw/XRT image 
(0.3-10 keV) of the same sky region, obtained by summing the two observations (obsID 00041799001 and 00041799002; see Table 3). Within the ISGRI error circle, the Seyfert AGN is detected as expected, consistent with its persistent X-ray nature. In contrast, source n.2 is not detected by 
\sw/XRT.  We have calculated a 3$\sigma$ upper limit to the XRT count rate of source n.2 at a level of  5.6$\times$10{$^{-3}$~counts~s$^{-1}$}, summing together these two \sw/XRT observations (0.3-10 keV) for a vignetting-corrected exposure time of 4.1 ks.
In Table~\ref{tab:xrtrates}  we summarize all results on \sw/XRT count rates and upper limits calculated using \textsc{sosta} in \textsc{ximage}, 
in the energy range 0.3-10 keV.

We note that the X-ray source n.2 is listed in the \xmm\ Serendipitous Source Catalog version 
4XMM-DR14 \citep{webb2020} under the designation 4XMM J164246.9+653553. Its X-ray coordinates are: 
R.A.=16$^\mathrm{h}$ 42$^\mathrm{m}$  46.92$^\mathrm{s}$,
Dec=+65$^{\circ}$ 35$\arcmin$ 53.87$\arcsec$ (J2000).  
We used the 63\% confidence level positional uncertainty provided in the catalog to compute a 90\% confidence error radius, which results in a value of 2\arcsec.2 (including both statistical and systematic uncertainties).

The EPIC observation 0670390401 is severely affected by background flares (Fig.~\ref{fig:xmmbgflares}). Therefore, we selected only the last $\sim$5~ks of the  exposure time, for any subsequent spectral and timing analysis.  
We extracted the source light curve of source n.2 in different energy ranges, finding no strong evidence of
hardness variability along the limited, clean exposure time. 
The EPIC spectra (net exposure time of 3.8~ks and 4.4 ks, respectively for the pn and each of the MOS cameras) 
is well fitted  by an absorbed power law model 
(C-stat = 200.38, 219 d.o.f; Fig. A.2), resulting into a photon index $\Gamma$=2.11$^{+0.70}_{-0.28}$, 
an absorbing column density N$_{\rm H}$$<0.12\times10^{22}$~cm$^{-2}$ and a flux corrected for the absorption 
UF=9.6 ($^{+3.1}_{-2.8}$) $\times10^{-14}$~erg~cm$^{-2}$~s$^{-1}$ (0.5-10 keV; Table~\ref{tab:xmmxrtspec}).

\subsubsection{Infrared and optical observations}
The arcsecond sized  positional accuracy of our proposed best candidate X-ray counterpart allow  us  to perform  a reliable search for counterparts at lower energies by using all the available catalogs in the HEASARC database. 

No archival infrared  source is located inside of it,  the nominal 2MASS limiting magnitude in the Ks band is $<$15.5 \citep{skrutskie2006}. Conversely, only one optical GAIA source  (listed in the catalog DR3 with ID 1635228086927497856) is found at a distance of 0.18\arcsec\  from  the X-ray positional centroid. Its magnitudes are G=19.9,  G$_{BP}$=19.9 and G$_{RP}$=19.5.   We note that GAIA mean G passband covers a wavelength range from the near ultraviolet (roughly 330 nm) to the near infrared (roughly 1050 nm) while the other two passbands, G$_{BP}$ and G$_{RP}$,  cover smaller wavelength ranges, from approximately 330 to 680 nm, and 630 to 1050 nm, respectively. The GAIA distance estimate, as obtained with the photogeometric method (which uses the parallax as well as the color and the apparent magnitude of the star), is  d=3.3$^{+3.2}_{-1.4}$ kpc \citep{bailer2021}. We note that this GAIA counterpart is also listed in the USNO-B1 catalogue \citep{monet2003} with magnitudes I=19.11, R2=19.77 and B2=19.76.


\begin{table}
\caption {\sw/XRT/PC net count rates (0.3-10 keV) and/or 3$\sigma$ upper limits to the count rates. Count rates and net exposure times are corrected for PSF, sampling dead time and vignetting.}
\label{tab:xrtrates} 
\scalebox{0.87}{
\begin{tabular}{cccc}
\hline
\hline    
source                     &     Obs ID              &      net exp              &        Count rate                 \\  
                                &                              &         (s)                             &        (count s$^{-1}$)                      \\  
\hline    
IGR J16426+6536   &   00041799001     &        2160                          &      $<$  6.76$\times$10$^{-3}$       \\
IGR J16426+6536   &   00041799002     &     1977                            &      $<$ 1.09$\times$10$^{-2}$      \\
IGR J16426+6536   &   41799001+ 41799002  &     4127                 &       $<$  5.64$\times$10$^{-3}$      \\       
\hline 
IGR J09446-2636   &   00047885001            &   1100      &        $<$2.10$\times$10$^{-2}$          \\
IGR J09446-2636   &   00041797001           &        600   &        $<$ 2.39$\times$10$^{-2}$        \\
IGR J09446-2636   &   00041797002           &   3871      &      (3.4$\pm{1.3}$) $\times$10$^{-3}$  \\
IGR J09446-2636   & 47885001+41797001 &        1700 &    $<$ 1.27$\times$10$^{-2}$            \\ 
\hline 
 IGR J21268+6203   &   00045405001        &    811       &         (1.29$\pm{0.58}$) $\times$10$^{-2}$  \\    
 IGR J21268+6203   &   00045405002        &   737       &  $<$ 2.49$\times$10$^{-2}$           \\
 IGR J21268+6203   &   00045405003        &  1294      &  $<$ 1.73$\times$10$^{-2}$           \\
 IGR J21268+6203   &   00045405004        &  468        & $<$  2.91$\times$10$^{-2}$           \\
 IGR J21268+6203   &   00045405005        & 1614       &  $<$   1.42 $\times$10$^{-2}$          \\
 IGR J21268+6203   &   00045405006        &  5066     &   (4.8$\pm{1.4}$) $\times$10$^{-3}$        \\
\hline 
IGR J02447+7046   &   00033613001     &        4486         &             (6.4$\pm{1.8}$ )$\times$10$^{-3}$           \\       
IGR J02447+7046   &   00032222004     &        237           &         $<$ 4.85$\times$10$^{-2}$                \\
IGR J02447+7046   &   00032222003     &      1437           &          $<$ 1.07$\times$10$^{-2}$          \\
IGR J02447+7046   &   00032222002     &        311           &          $<$ 2.28$\times$10$^{-2}$           \\   
IGR J02447+7046   &   00032222005     &       3580           &             (6.3$\pm{1.5}$) $\times$10$^{-3}$        \\
IGR J02447+7046   &   32222002+3+4   &       1982          &       $<$ 9.38$\times$10$^{-3}$           \\
\hline
\hline  
\end{tabular}}
\end{table} 


\subsection{IGR J09446$-$2636}

\subsubsection{Hard X-ray observations}

Our investigation with the bursticity method of the  15 years IBIS/ISGRI data archive revealed that IGR J09446$-$2636 has been detected in outburst only once. The source is best detected in the softer band 17-30 keV ($\sim$5$\sigma$, effective exposure of $\sim$7 ks) during the temporal period from  6 May 2005 20:10 UTC  to 7 May 2005 11:31 UTC in revolution 313. The measured average X-ray flux is 6.2$\pm$1.3 mCrab (17-30 keV). We checked that the source was not detected when it was in the satellite FoV during the previous revolution 312  from 3 May 2005 19:15 UTC to 6 May 2005 09:15 UTC,  the derived 3$\sigma$ flux upper limit is 3.6 mCrab (17-30 keV, effective exposure of 10 ks). Similarly,  the source was not detected in the 
period subsequent to the outburst detection, from 7 May 2005 12:00 to 9 May 2005 09:10, and the derived  
3$\sigma$ flux upper limit is  3.4 mCrab (17-30 keV, effective exposure of 11 ks). This allows us to firmly constrain the outburst duration to   $\sim$0.6 days. 

The average IBIS/ISGRI spectrum of the source during the outburst is best fit with a power law model ($\chi^{2}_{\nu}$=0.9, 2 d.o.f.)  with a soft photon index $\Gamma$=4.8$^{+2.8}_{-2.2}$. The average 17-30 keV (20--40 keV) flux is 3.8$\times$10$^{-11}$ erg cm$^{-2}$ s$^{-1}$ (2.6$\times$10$^{-11}$ erg cm$^{-2}$ s$^{-1}$). A good fit is also achieved with a blackbody model ($\chi^{2}_{\nu}$=0.7, 2 d.o.f.)  with temperature kT=3.5$^{+2.2}_{-1.5}$ keV or alternatively by a thermal bremsstrahlung model ($\chi^{2}_{\nu}$=0.8, 2 d.o.f.)  with temperature kT=7$^{+14}_{-4}$ keV.

\subsubsection{Soft X-ray  observations}

\begin{figure}
\begin{center}
\includegraphics[height=9 cm, angle=0]{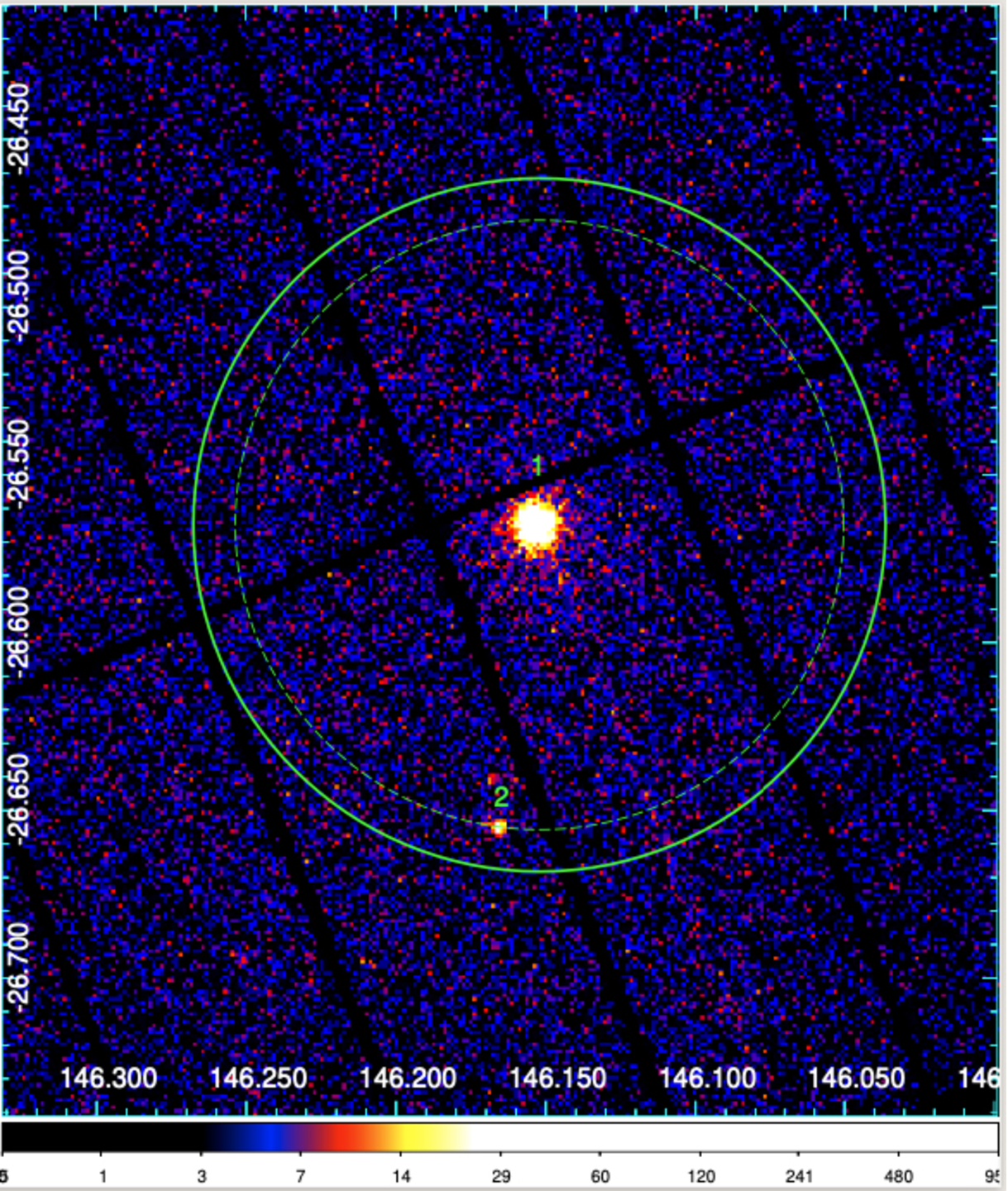}
\caption{\xmm\ EPIC pn  image (0.2-10 keV) of IGR J09446$-$2636 sky region 
with superimposed the ISGRI  error circle at 90\% confidence (dashed) and at 95\% confidence. Two X-ray sources (n.1 and n.2) are detected inside of it.} 
\end{center}
\end{figure} 

Fig.~2  shows the \xmm/EPIC pn image of IGR J09446$-$2636 sky region as obtained from obsid 0761112901  in Table 2. The image is overlaid with the ISGRI error circles. Thanks to this X-ray observation, we can investigate it in the X-ray band in order to search for the best candidate  X-ray counterpart. The X-ray source n.1 is the Seyfert AGN  previously proposed in the literature as counterpart. We note that such X-ray source  has been also  detected with \sw/XRT during all three available observations (see Table 3), this is expected given the X-ray persistent nature of Seyfert AGN.  Notably, there is an additional  detected source in the \xmm\ image, i.e. n.2, which has never been considered as a potential counterpart to date.

This \xmm\ observation is severely affected by high background flares, compelling us to select only the time interval with the lowest background level, lasting about 3~ks (enclosed by the two vertical dashed lines in Fig.~\ref{fig:xmmbgflares-igr09446}).
The EPIC pn, MOS~1 and MOS~2 spectra of source n.2 are soft, i.e. almost all counts are concentrated below 2 keV. In fact, a fit with an absorbed power law model results into a photon index, $\Gamma$, of 9. Therefore we adopted an emission model from hot plasma (\texttt{apec} in \texttt{XSPEC}), assuming solar abundances. 
If we take into account the fact that the Galactic absorption towards the source is N$_{H}$=6.5$\times$10$^{20}$ cm$^{-2}$, we can fit the EPIC spectra fixing the absorption to this value (see Fig. A.4). We obtain the following parameters: 
kT=0.37$^{+0.19}_{-0.07}$ keV and a flux corrected for the absorption
UF=8.6$^{+1.9}_{-1.8}$~$\times$10$^{-14}$ erg cm$^{-2}$ s$^{-1}$ (0.5-10 keV; Table~\ref{tab:xmmxrtspec}).


The spectroscopy of the \sw/XRT spectrum extracted from the observation with the longest exposure (ObsID 00041797002) results into source parameters fully consistent with the EPIC results.
We report them in Table~\ref{tab:xmmxrtspec}.

The X-ray source n.2  has not been detected in the \sw/XRT summed image of observations 00047885001 and 00041797001 (results in Table~\ref{tab:xrtrates}).
We inferred a 3$\sigma$ upper limit on the XRT/PC rate of 1.27$\times$10$^{-2}$ counts s$^{-1}$ (0.3-10 keV), which translates into an upper limit to the flux corrected for the absorption UF=3.32$\times$10$^{-13}$ erg cm$^{-2}$ s$^{-1}$ (0.5-10 keV), assuming the
same spectral parameters observed before (i.e. N$_{H}$=6.5$\times$10$^{20}$ cm$^{-2}$ and a hot plasma temperature of 0.39 keV (\texttt{apec} in \texttt{WebPIMMS}). We note that this flux is not constraining 
with respect to the source spectra extracted from longer exposure observations reported here.

The X-ray source n.2 is reported in the \xmm\ serendipitous source catalog 
(version XMM-DR14) as 4XMM J094439.8-263919. Its X-ray coordinates are  R.A.=09$^\mathrm{h}$ 44$^\mathrm{m}$ 39$^\mathrm{s}$.86, 
Dec.=$-26^{\circ}$ 39$\arcmin$ 19$\arcsec$.01 (J2000).  We used the 63\% confidence level positional uncertainty provided in the catalog to compute a 90\% confidence error radius, which results in a value of 4.\arcsec1  (including both statistical and systematic uncertainties).

\subsubsection{Infrared and optical observations}
We used the arcsecond sized  positional accuracy of our proposed best candidate X-ray counterpart of IGR J09446$-$2636  to perform  a search for counterparts at lower energies by using all the available catalogs in the HEASARC database. 

One archival 2MASS near-infrared  source is located inside of it, its magnitudes are J=11.96, H=11.34 and K=11.12, respectively. If we use the reddening-free NIR diagnostic $Q$ of \cite{neg2007}  then we find a  $Q$ value of 0.24 which is typical of late-type stars  (K or M). 
This infrared counterpart has been also detected by GAIA  (catalog DR3, ID 5658291556154316544).
%
Two  distance estimates are available as obtained with two different methods: geometric (based only on the parallax) and photogeometric (which also uses the colour and the apparent magnitude of the star). Both methods provide  the same distance  value, hence we assume a distance of  108.94$\pm$0.25  pc (Bailer-Jones et al. 2021).  In addition, we note that the GAIA counterpart is also listed in the USNO-B1 catalogue with magnitudes I=13.03, R1=14.76 and B1=16.87.


\subsection{IGR J21268$+$6203}

\subsubsection{Hard X-ray  observations}

IGR J21268$+$6203 has been detected in outburst only once as revealed by our investigation with the bursticity method of the  15 years IBIS/ISGRI data archive.  The source is detected in the 
energy band 20-100 keV ($\sim$6$\sigma$, effective exposure of $\sim$33 ks) during revolution 511 from  19 December 2006 21:07 UTC  to 22 December 2006 06:28 UTC.  A similar significant detection  is also obtained in other energy bands (e.g. $\sim$5$\sigma$ in 18-60 keV band). In order to firmly constrain the duration of the outburst activity, which is $\sim$ 3 days,  we checked that the source was not significantly detected  by ISGRI when it was in the satellite FoV during the previous and the subsequent revolution (510 and 512, respectively).
 
The IBIS/ISGRI spectrum  during the outburst is fit with a power law 
model ($\chi^{2}_{\nu}$=0.4, 6 d.o.f.) with photon index $\Gamma$=0.8$^{+0.8}_{-0.9}$. The average 18-60 keV flux is 2.9$\times$10$^{-11}$ erg cm$^{-2}$ s$^{-1}$.

\begin{figure}
\begin{center}
\includegraphics[height=8 cm, angle=0]{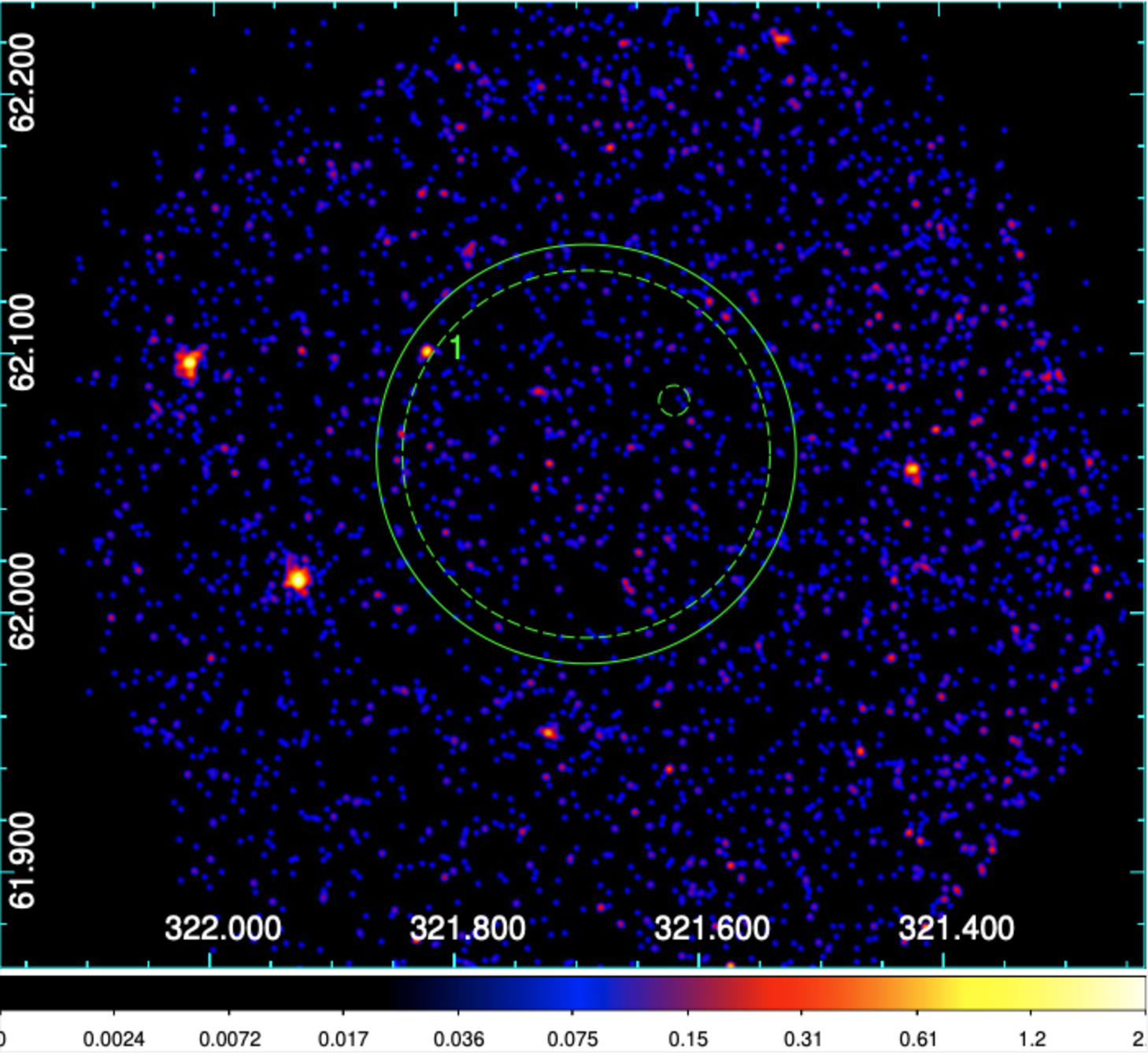}
\caption{\sw/XRT image (0.2-10 keV) of IGR J21268+6203 sky region with superimposed the ISGRI  error circle at 90\% (dashed) and 95\% confidence. 
Only one X-ray counterpart (n.1) is detected inside of it. The smallest dashed circle marks the position of the radio/infrared counterpart previously proposed in the literature, which is undetected by \sw/XRT.} 
\end{center}
\end{figure} 

\subsubsection{Soft X-ray  observations}

Fig. 3 shows the  \sw/XRT image (0.2-10 keV) of IGR~J21268+6203 sky region as obtained from the sum of all six available observations  (to increase the statistics) listed in Table~\ref{tab:xrtrates}. This allows us to investigate in the X-ray band the entire ISGRI error circle, in order to search for the best candidate  X-ray counterpart. We note that the  radio/infrared source, candidate AGN previously proposed as counterpart of IGR~J21268+6203, 
is not detected  in X–rays. This casts doubts on its associations with the INTEGRAL transient. 
Conversely, only one X-ray source (n.2) is detected, which we consider as best candidate X-ray counterpart. 
We obtained the best determined XRT position 
at R.A. (J2000) = $21^{\rm h}27^{\rm m}17\fs71$ Dec (J2000) = $62^\circ06\arcmin04\farcs5$ with a 90$\%$ confidence error radius of 4$''$.3  by using the UK Swift Science Data Centre tools\footnote{http:/www.swift.ac.uk/user$\textunderscore$objects}. 

We then extracted the \sw/XRT spectrum from the observation with the longest exposure time (ObsID 00045405006), where the X-ray source n.2 is detected with a count rate of (4.8$\pm{1.4}$)$\times$10$^{-3}$~count~s$^{-1}$ (0.3-10 keV).
Almost all its net counts are below 2 keV. In fact, if  fitted with an absorbed power law model,   a photon index of $\sim$9 is found.
Therefore, a hot plasma model (\textsc{apec} in  \textsc{xspec}) 
is assumed in the spectral fitting. Moreover, since the X-ray source is associated with a nearby star 
(details in the next section) with an estimated absorbing column density
of N$_H$=$6\times$10$^{20}$\,cm$^{-2}$ (according to the
tool 3D-N$_H$ \footnote{http://astro.uni-tuebingen.de/nh3d/nhtool}), 
we fixed the N$_H$ to this value.
The results are reported in Table~\ref{tab:xmmxrtspec}
and shown in Fig.~A.5. 

\subsubsection{Infrared and optical observations}

We searched for lower energy counterparts of IGR~J21268+6203 by using all the available catalogs in the HEASARC database.
Within the arcsecond sized error circle of the best X-ray candidate counterpart, only one bright 2MASS infrared source is found 
(21271790+6206035).  
Its magnitudes are J=4.96, H=4.48, and K=4.32. By applying the reddening-free near-infrared diagnostic parameter $Q$ introduced by Negueruela \& Schurch (2007), we obtain a value of $Q$=0.2, which is typical of late-type K-M stars. Indeed, the infrared counterpart is identified in the SAO optical catalog \citep{whipple1966} as a K0-type star with a visual magnitude of V=7.6. Furthermore, it is listed in the USNO-B1 catalog  with magnitudes I=5.86, R1=6.52, and B1=8.64. The GAIA DR3 catalog provides a precise distance estimate for this object (Gaia ID 2192883518206014848), offering two values derived from independent methods. Both yield a consistent distance value, so  we assume a distance of 272.2$\pm$1.3 pc (Bailer-Jones et al. 2021).

\begin{table*}
\caption {Spectroscopy of the \xmm\ and \sw/XRT observations}
\label{tab:xmmxrtspec} 
\begin{tabular}{ccccccrc}
\hline
\hline    
source         &     Obs ID  &  N$_H$                & $\Gamma$ &    kT     &       UF (0.5-10 keV)                 &  C-stat (dof)  &  L$_X$ (0.5-10 keV)   \\  
               &             & (10$^{22}$ cm$^{-2}$) &          &  (keV)    &     (erg cm$^{-2}$ s$^{-1}$)          &                &   (erg s$^{-1}$)     \\     
\hline 
IGRJ16426+6536 & 0670390401  & $<$0.12       & 2.11$^{+0.70}_{-0.28}$ &  $-$  & 9.6$^{+3.1}_{-2.8}$$\times$10$^{-14}$ &  200.38 (219) &  1.25$\times$10$^{32}$   \\  
\hline
IGRJ09446-2636  & 00041797002  & 0.065 (fixed)          & $-$ & 0.39$^{+0.48}_{-0.17}$ & 8.7$^{+5.3}_{-4.3}$$\times$10$^{-14}$   & 8.96 (16) & 1.2$\times$10$^{29}$ \\  
IGRJ09446-2636  & 0761112901  & 0.065 (fixed)          & $-$ & 0.37$^{+0.19}_{-0.07}$ & 8.6$^{+1.9}_{-1.8}$$\times$10$^{-14}$ & 154.19 (201) & 1.2$\times$10$^{29}$   \\  
\hline  
IGRJ21268+6203 & 00045405006  & 0.06 (fixed) &  $-$          & 0.54$^{+0.22}_{-0.28}$ & 5.4$^{+3.5}_{-2.6}$$\times$10$^{-14}$ &   7.82 (9)  & 4.8$\times$10$^{29}$   \\ 
\hline 
IGRJ02447+7046 & 00033613001 & $<$10         & 0.8$^{+3.6}_{-1.1}$    &  $-$  & 4.8$^{+5.2}_{-2.3}$$\times$10$^{-13}$ &   14.50 (20)  &  2.8$\times$10$^{32}$   \\ 			      
               & 00033613001 & 0.6 (fixed)   & 0.76$^{+0.84}_{-0.85}$ &  $-$  & 4.8$^{+5.2}_{-2.3}$$\times$10$^{-13}$ &   14.51 (21)  &  2.8$\times$10$^{32}$   \\                                                    
IGRJ02447+7046 & 00032222005 & $<$5.2        & 1.6$^{+2.7}_{-1.8}$    &  $-$  & 4.1$^{+60.}_{-2.1}$$\times$10$^{-13}$ &   3.10 (12)   &  2.4$\times$10$^{32}$   \\ 
               & 00032222005 & 0.6 (fixed)   & 1.1$^{+1.0}_{-1.0}$    &  $-$  & 3.9$^{+60.}_{-1.9}$$\times$10$^{-13}$ &   3.26 (13)   &  2.2$\times$10$^{32}$   \\  
\hline
\hline  
\end{tabular}
\end{table*}

\subsection{IGR J02447+7046}

\subsubsection{Hard X-ray observations}

From our analysis of the 15 years IBIS/ISGRI data archive  using the bursticity method, we found that IGR J02447+7046 was detected in outburst only once. The source is best detected in the energy band 18-60 keV at 5.2$\sigma$ level, a slighgtly less significant detection is also achieved in the band 20-40 keV. The  measured average X-ray flux is 29.6$\pm$5.9 mCrab (18-60 keV). The onset of the hard X-ray activity is on 12 November 2003 16:20 UTC (rev 132), unfortunately the IBIS/ISGRI temporal coverage of the outburst is limited, as the source moved out of the FoV after 17 November 2003. When it re-entered the FoV about one months later, on 12 December 2003 (rev 142), it was no longer active, as no significant detection was achieved in the 18-60 keV band. Therefore the duration of the outburst can be constrained in the range 5-30 days, contrary to a previous report in the literature that estimated it as only $\sim$3 days (Bird et al. 2016). 

We extracted the average IBIS/ISGRI spectrum of the source during the outburst, which is fit by using a  power law model ($\chi^{2}_{\nu}$=0.25, 3 d.o.f.)  with photon index $\Gamma$=2.4$\pm$2.0.
The average 18-60 keV  flux is $\sim$1.0$\times$10$^{-10}$ erg cm$^{-2}$ s$^{-1}$.

\subsubsection{Soft X-ray observations}

\begin{figure}
\begin{center}
\label{fig:ima_igr02447}
\includegraphics[height=8.4 cm, angle=0]{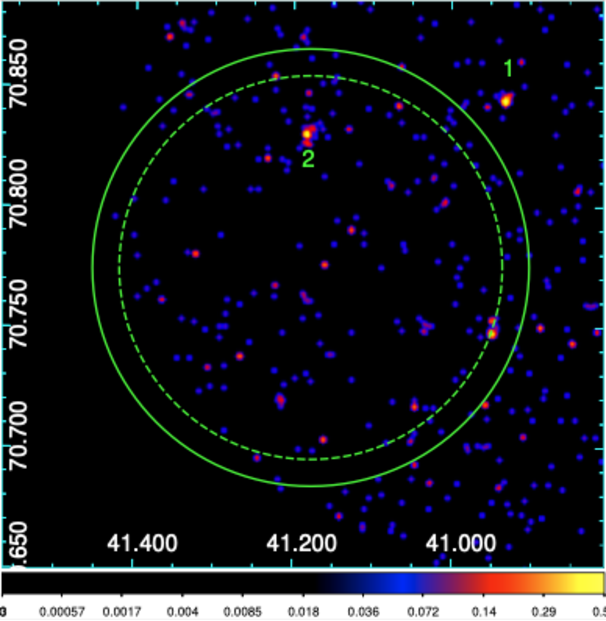}
\caption{\sw/XRT image (0.2-10 keV, obsid 00033613001 in Table 3) of IGR J02447+7046 sky region with superimposed its IBIS/IGRI  error circle  at 90\% confidence (4$\arcmin$.78 radius, dashed) and at 95\% confidence (5$\arcmin$.45 radius). Only one X-ray counterpart (n.2) is  detected inside the 95\% confidence error circle. The detected X-ray source n.1, outside the   95\% confidence error circle, is the AGN previously suggested in the literature as counterpart of IGR J02447+7046} 
\end{center}
\end{figure}

Table 3 lists five \sw/XRT observations of IGR J02447+7046 which can be used  
to search for its  best candidate X-ray counterpart. 
 
Fig. 4 shows the  \sw/XRT image of the source sky region by using the observation with the longest exposure time (obsid 00033613001 in Table 3). The X-ray source n.1 is the Seyfert AGN previously proposed in the literature as  counterpart of IGR J02447+7046. We note that such X-ray source has been also detected by \sw/XRT in all the remaining analyzed observations (obsid 00032222005 and sum of 32222002+3+4), as  expected given its X-ray persistent nature. Strikingly, the AGN is located outside the ISGRI large error circle at 95\% confidence (5$\arcmin$.45 radius). Conversely, there is  only one detected X-ray source inside of it (n.2), which  has never been considered as a potential counterpart to date. In Fig. 4 we note that the  95\% INTEGRAL error circle is not entirely covered by the FoV of the \sw/XRT observation. However, only a very small rim  
(whose area is $\sim$3\% of the total error circle area) remains outside the FoV. This uncovered portion is therefore very negligible, and it has no practical impact on counterpart identification: the 97\% of the total error circle area  is imaged by \sw/XRT and this  ensures that the search for the correct X-ray counterpart is not significantly affected. The identification of the proposed counterpart remains robust. 

We performed the spectroscopy of source n.2 from ObsID 33613001 (performed in 2015) and from ObsID 32222005 (in 2016).  
At first we let  the N$_H$ parameter free during the fit. Then we fixed it to the column density towards the likely optical counterpart (details in the next section), resulting from 
the online tool 3D-N$_H$ \footnote{http://astro.uni-tuebingen.de/nh3d/nhtool}, which returns N$_H$=$(0.6\pm{0.3})$$\times$10$^{22}$\,cm$^{-2}$. 
The results obtained with both fits are reported  in Table~\ref{tab:xmmxrtspec}.  

We note that three  \sw/XRT observations have been performed in February 2012 in a short time interval of a few days, hence they have been summed up (obsid 32222002+3+4) to increase the exposure time and the statistics (vignetting corrected exposure of 1982 seconds; see Table~\ref{tab:xrtrates}).
Source n.2 was not detected and we estimated a 3$\sigma$ upper limit on the net count rate of 9.38$\times$10$^{-3}$ count  s$^{-1}$ (0.3-10 keV; see Table 4). 
Assuming a power law model with $\Gamma$=1 and N$_H$$<$0.6$\times$10$^{22}$\,cm$^{-2}$ (see Table~\ref{tab:xmmxrtspec} for source spectra taken during the other XRT observations),
this count rate translates into an un-absorbed flux UF=8.5$\times$10$^{-13}$ erg cm$^{-2}$ s$^{-1}$ (0.5-10 keV).
 
We obtained the best determined \sw/XRT position of X-ray source n.2 
at R.A.(J2000)=$02^{\rm h}44^{\rm m}43\fs20$ Dec(J2000)=$70^\circ49\arcmin47\farcs7$ with a 90$\%$ confidence error radius of 4$''$.8  by using the UK Swift Science Data Centre tools.

\subsubsection{Infrared, optical and radio observations}

\begin{figure}
\begin{center}
\includegraphics[height=8 cm, angle=0]{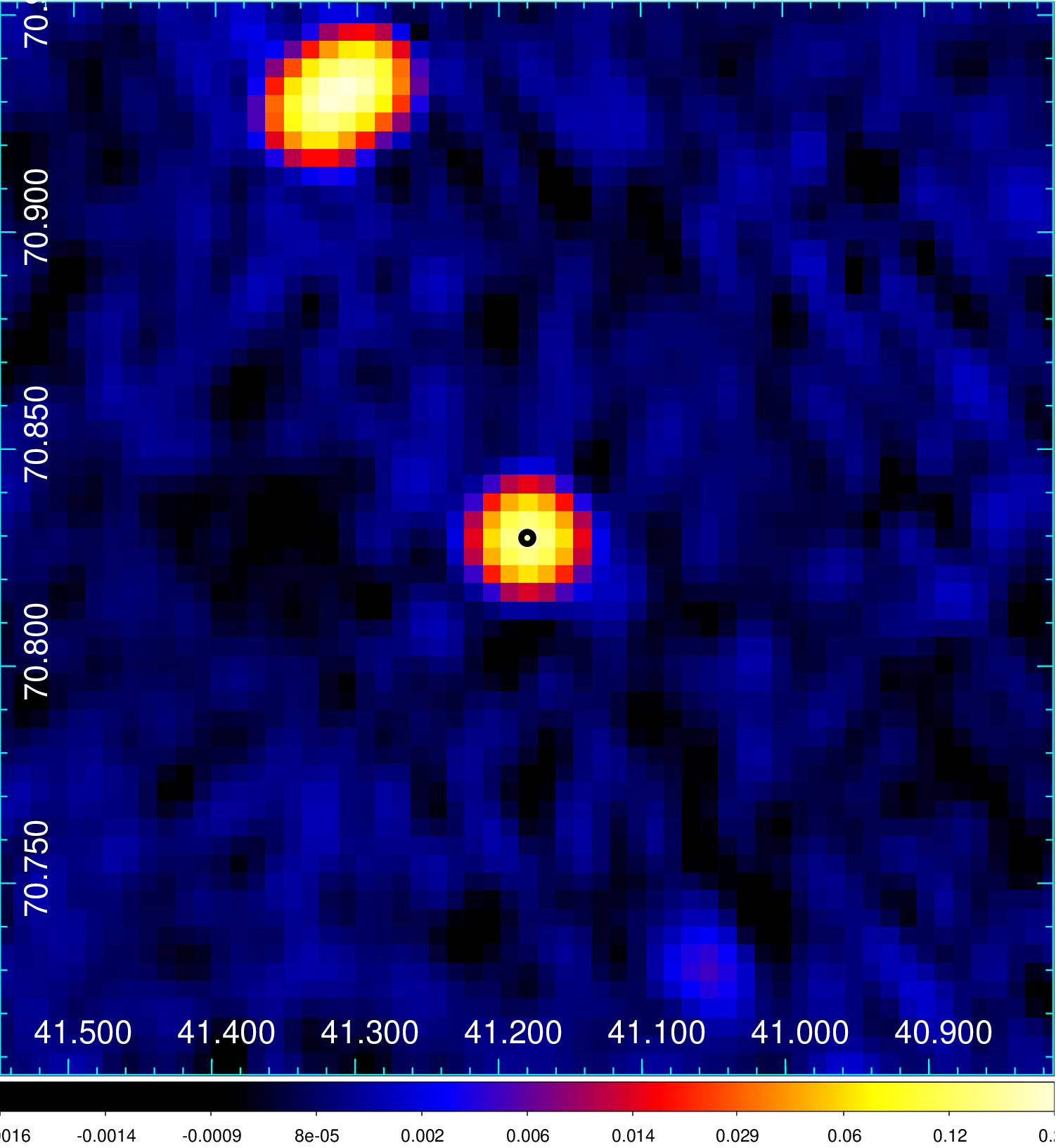}
\caption{NVSS radio map at 20 cm (as taken from the NVSS team web page http://www.cv.nrao.edu/nvss/) of IGR J02447+7046 sky region. The small black circle marks the position of the \sw/XRT error circle of the best candidate X-ray counterpart of IGR J02447+7046 (source n.2 in Fig.~4)}
\end{center}
\end{figure}

We used the  arcsecond sized  positional accuracy of the best candidate X-ray counterpart 
to  search for counterparts at lower energies by using all the available catalogs in the HEASARC database. 
Only one 2MASS infrared  source is located inside of it, with magnitudes  J=14.40, H=13.98 and K=13.74. 
If we use the reddening-free NIR diagnostic $Q$ of Negueruela \& Schurch (2007) then we find a  $Q$ value of 0.02 which is typical of early-type stars (O or B). This infrared counterpart has been also detected in the optical band as listed in both catalogs GAIA DR3 (545121204876403200) and USNO-B1, respectively. The USNO optical magnitudes are R1=15.90 and B1=18.02, respectively. Two equal distance estimates are available as obtained by GAIA with two different methods (geometric and photogeometric), hence we assume a distance of 2.2$^{+0.3}_{-0.2}$ kpc (Bailer-Jones et al. 2021).  

Notably, the  radio source  NVSS J024443+704946 is located inside the \sw/XRT error circle, its position is fully consistent with the position of the pinpointed optical/infrared 
counterpart.  It is characterized by a 20 cm radio flux  of 205 mJy, it is also detected at longer radio  wavelength  with flux values equal to 474 mJy (92 cm, Westerbork Synthesis Radio Telescope) and 
707 mJy (200 cm, Giant Metrewave Radio Telescope).  Fig. 5 shows the \sw/XRT error circle of source n.2 superimposed on the NVSS 20 cm radio map of the sky region. It is evident the spatial match with the bright radio source  NVSS J024443+704946. We considered the  possibility that such association could be  spurious, and to this aim we calculated the  probability  of  finding  a  NVSS radio source  inside the \sw/XRT error circle by chance. The probability that the nearest unrelated NVSS source lies within angular distance r of any position is $P$ = 1 - e$^{-\pi\rho r^2}$   (where r is the distance between the XRT centroid and the associated NVSS source, and $\rho$  is the  spatial density of NVSS sources). Given that the NVSS catalog contains $\sim$50 sources per square degree \citep{condon1998}, we estimated a chance coincidence of 2.6$\times$10$^{-4}$ (i.e. a probability of 0.026\%).  Such significantly  low probability  
supports a real  physical association.


\section{Discussion}
We discuss the overall results about the newly proposed X-ray counterparts 
of the four INTEGRAL  sources in the following sub-sections.

\subsection{IGR J16426+6536}

IGR J16426$+$6536 is a hard X-ray transient detected in outburst only once (June 2004) 
by investigating the 15-year IBIS/ISGRI archive. The outburst duration is constrained to 1-150 days, as only the first day of the rising X-ray activity was covered by INTEGRAL before the source moved out of its FoV. The measured average 18–60 keV flux (luminosity at 3.3 kpc) is $\sim$5.5$\times$10$^{-11}$ erg cm$^{-2}$ s$^{-1}$ ($\sim$7$\times$10$^{34}$ erg s$^{-1}$), though the source likely became brighter during the unobserved phase. 
Unfortunately, no all-sky soft or hard X-ray monitors (like MAXI and \sw/BAT, respectively) were available at the time to track a potential brighter phase. Despite extensive INTEGRAL coverage 
(effective exposure $\sim$2.7 Ms), no persistent hard X-ray emission was detected. We derive a 20–40 keV 3$\sigma$ upper limit flux (luminosity) of 0.3 mCrab or 2.3$\times$10$^{-12}$ erg cm$^{-2}$ s$^{-1}$ (3$\times$10$^{33}$ erg s$^{-1}$). 

We used archival \xmm\  and \sw/XRT soft X-ray data to perform a new analysis of the entire INTEGRAL large error circle. Two X-ray sources were detected: source n.1 is X-ray persistent and a confirmed Seyfert AGN; source n.2 is a weak and unidentified X-ray emitter.  
Previously, the AGN had been proposed as counterpart of IGR J16426+6536 based on circumstantial evidence. Here we note that this interpretation conflicts with its transient hard X-ray nature  as well as with the INTEGRAL upper limit on its persistent X-ray emission, as all firmly identified Seyfert AGNs detected by INTEGRAL are persistent  X-ray sources. Furthermore, the AGN redshift value (z=0.323) is significantly  higher than typical values of INTEGRAL Seyfert AGNs (i.e. average z=0.033, maximum z=0.1; Butler et al. 2009), clearly IGR J16426+6536 would be an outlier and this could be  very difficult to justify. In addition, IGR J16426$+$6536 is not listed in the \sw/BAT 157-months survey hard X-ray catalog \citep{lien2025} which reports almost 1,000 firmly identified Seyfert AGNs, further supporting the evidence that  it is not  a such object. Likely it is too distant to be detected at hard X-ray by INTEGRAL or \sw/BAT. In fact, a simple flux–luminosity argument indicates that an AGN at the redshift of XMMSL1 J164303.7+653253 (z=0.323) would be far too faint to be detected in hard X-rays by INTEGRAL or \sw/BAT. At this redshift the expected hard X-ray flux, even for a X-ray luminosity of $\sim$10$^{44}$ erg  s$^{-1}$ (20-40 keV), would be of the order of $\sim$ 3$\times$10$^{-13}$ erg cm$^{-2}$ s$^{-1}$, well below the sensitivity of both INTEGRAL/IBIS and Swift/BAT. At their typical surveys depth only X-ray sources with L$_x$$>$4$\times$10$^{45}$ erg  s$^{-1}$ and L$_x$$>$2$\times$10$^{45}$ erg  s$^{-1}$ could be detected at z=0.323 by INTEGRAL and \sw/BAT, respectively. These values exceed by more than an order of magnitude the hard X-ray luminosities typically measured for the bulk of the local Seyfert population detected in INTEGRAL  and BAT surveys, whose median redshift is z$\sim$0.03 (Butler et al. 2009, Ajello et al. 2012).

One possible classification for the AGN  XMMSL1 J164303.7+653253 is that of a Narrow-Line Seyfert 1 (NLS1). While a handful of NLS1s have indeed been detected as $\gamma$-ray emitters with Fermi-LAT \citep{abdo2009}, these constitute a rare  subgroup, generally associated with radio-loud systems hosting relativistic jets \citep{yuan2008}. The vast majority of NLS1s, including those identified in hard X-ray surveys, show no evidence of $\gamma$-ray emission above 100 MeV \citep{ajello2012}. Therefore, while the NLS1 classification would in principle make a $\gamma$-ray emitting scenario physically possible, the absence of any $\gamma$-ray detection for XMMSL1 J164303.7+653253 is not surprising and does not necessarily support the association with IGR J16426+6536.

Taken at face value, all these findings suggest that the Seyfert AGN is not the true counterpart of IGR J16426+6536. Conversely, the X-ray characteristics of source n.2 are  more compatible with those of IGR J16426$+$6536. Hence we propose it as the best candidate X-ray counterpart. Within its arcsecond-sized error circle, we pinpointed a single optical object at $\sim$3.3 kpc, placing it within the boundaries of our Galaxy. Although the relatively high Galactic latitude of IGR~J16426+6536 (b$\sim$38$^{\circ}$) could initially suggest an extragalactic origin, such position is fully consistent with a Galactic object located above the Galactic plane at a vertical height of z$\sim$2 kpc. 

Based on our results, we propose that IGR J16426$+$6536 is a Galactic X-ray binary. His elevated  vertical height  above the Galactic plane is unusual for disk sources, which typically have z$<$1 kpc,  but remains compatible with low-mass X-ray binaries (LMXBs) or Very Faint X-ray Transients (VFXTs) that have received strong natal kicks or are formed in the Galactic halo, e.g. the LMXB XTE J1118+480 with z$>$1.5 kpc (\cite{jonker2004}; \cite{repetto2012}). In this context, the magnitude values of the optical counterpart as well as the near-infrared magnitude upper limit 
support a faint low-mass donor star for IGR J16426$+$6536, hence favouring a classification as a  LMXB or a VFXT in the halo.
The transient X-ray nature of IGR J16426$+$6536, its outburst lasting 1-150 days with lower limit X-ray luminosity of  $\sim$7$\times$10$^{34}$ erg s$^{-1}$,  its quiescent X-ray state of 1.2$\times$10$^{32}$ erg $^{-1}$, all are consistent with the typical behavior of VFXTs (\cite{Wijnands2006}, \cite{degenaar2009}), like other such systems detected by INTEGRAL \citep{sgu2024}.  
Alternatively, the source could be a classical LMXB if we consider the eventuality  of a potentially unobserved brighter X-ray phase.


\subsection{IGR J09446$-$2636}

IGR J09446$-$2636 was detected by INTEGRAL in outburst only once (May 2005), by investigating the 15-year IBIS/ISGRI archive. The outburst duration is $\sim$0.6 days, revealing a short hard X-ray transient nature. The source is best detected in the 17–30 keV band, with a soft power law spectrum ($\Gamma$$\sim$4.8) indicative of thermal emission; indeed a good fit is also obtained with a blackbody or bremsstrahlung model. The average 17–30 keV flux (luminosity at 109 pc) is $\sim$3.8$\times$10$^{-11}$ erg cm$^{-2}$ s$^{-1}$ ($\sim$5.5$\times$10$^{31}$ erg s$^{-1}$). Despite significant INTEGRAL coverage (effective exposure $\sim$560 ks), no persistent hard X-ray emission was detected. We derive a 20–40 keV 3$\sigma$ upper limit flux (luminosity) of 0.9 mCrab or 6.9$\times$10$^{-12}$ erg cm$^{-2}$ s$^{-1}$ (9.3$\times$10$^{30}$ erg s$^{-1}$).

Using archival  \xmm\ and \sw\  X-ray observations, we investigated the entire INTEGRAL error circle in order to search for the soft X-ray counterpart. Two X-ray sources were detected
inside of it: source n.1 is X-ray persistent and a confirmed Seyfert AGN; source n.2 is a weak X-ray emitter and unidentified. Although the AGN was previously proposed in the literature as counterpart, here we note that its persistent X-ray nature is inconsistent with the transient X-ray behavior of 
IGR J09446$-$2636 as well as  its INTEGRAL upper limit on the persistent X-ray emission. Furthermore, the ISGRI spectrum of the source is soft and thermal, inconsistent with an AGN nature. We note that the AGN redshift (z=0.14)  exceeds the typical values of Seyfert AGNs firmly detected by INTEGRAL (average z=0.033), this would make IGR J09446$-$2636 an outlinear and it could  be  very difficult to explain it. In addition,  IGR J09446-2636 is not listed in the \sw/BAT 157-months survey hard X-ray catalog, which reports almost 1,000 firmly identified Seyfert AGN, further supporting  the evidence that  it is  not an AGN. Conversely, the X-ray characteristics of source n.2 are more compatible with those of 
IGR J09446-2636. In fact, its X-ray spectrum  is soft and thermal as measured by \xmm, this fully agrees with the soft spectrum measured above  17 keV by ISGRI. Hence, we propose the X-ray source n.2  as the true  X-ray counterpart of IGR~J09446-2636. 

Within its arcsecond-sized error circle, we identified a single optical/infrared object at $\sim$109 pc, placing it well within the boundaries of our Galaxy. Although the relatively high Galactic latitude of
IGR~J09446$-$2636 (b$\sim$20$^{\circ}$) could suggest an extragalactic origin, this position is fully consistent with a very nearby Galactic object located  above the Galactic plane with vertical height z$\sim$38 pc. Its photometric properties are consistent with a late-type star (F to M spectral type).

The ratio of X-ray to optical flux can be used to identify its nature \citep{Maccacaro1988}: the optical counterpart has a V=15.8 mag \citep{Zacharias2013}, the observed flux in the 0.3-3.5 keV band is 8$\times$10$^{-14}$ erg cm$^{-2}$ s$^{-1}$, leading to 
log$(\frac{f_X}{f_{\mathrm{opt}}})=-1.4$, consistent with coronal emission from M or K-type stars.

In the light of our reported results on  IGR J09446$-$2636, we propose its identification as a flaring star, a class of active late-type objects known to produce short and energetic outbursts both in the soft and hard X-ray band (\cite{gudel2004}; \cite{maggio2008}). Such stars are commonly found at intermediate Galactic latitudes due to their proximity and to the vertical scale height of the local stellar population \citep{fleming1995}. The X-ray emission arises from magnetic energy release in the stellar coronae of magnetically active stars, such as solar-type stars, K-M dwarfs, or pre-main-sequence stars. Stellar X-ray flares typically last from hours to days, with peak X-ray luminosities in the range 10$^{29}$–10$^{34}$ erg s$^{-1}$ \citep{zhao2024}. In particular, hard X-ray emission from flaring stars is mainly thermal in origin, being best interpreted by single or multi temperature models of an optically thin plasma. The soft and thermal  X-ray spectrum  measured with both \xmm\ and ISGRI is in agreement with such interpretation. Moreover, the flare duration and its hard X-ray luminosity match expectations for stellar flares detected by INTEGRAL in the hard X-ray band \citep{sgu2016}.


\subsection{IGR J21268+6203}

From our analysis, we find that IGR J21268+6203 is a hard X-ray transient, detected in outburst only once throughout the 15-year IBIS/ISGRI archival dataset. The outburst duration is constrained to $\sim$3 days, with an average 18–60 keV flux (luminosity) of $\sim$2.9$\times$10$^{-11}$ erg cm$^{-2}$ s$^{-1}$
($\sim$2.5$\times$10$^{32}$ erg  s$^{-1}$). Despite the extensive INTEGRAL coverage (effective exposure of $\sim$3.3 Ms), no persistent hard X-ray emission was detected. We derive a 3$\sigma$ upper limit on the 20–40 keV flux of 0.3 mCrab or  
2.3$\times$10$^{-12}$ erg cm$^{-2}$ s$^{-1}$. 

To identify the most likely soft X-ray counterpart, we analyzed \sw/XRT archival observations, fully covering the INTEGRAL error circle. A single X-ray source was detected within it, which we propose as the best candidate X-ray counterpart.
Within its arcsecond-sized error circle, we identified a single bright optical/infrared source, classified as a K0-type star located at a distance of $\sim$272 pc. Its Galactic latitude of 
$\sim$8 $^{\circ}$  is fully consistent with a nearby Galactic object located slightly above the Galactic plane (z$\sim$38 pc). 

IGR J21268+6203 had previously been proposed in the literature as a candidate AGN, primarily based on its position slightly off the Galactic plane as well as  on the presence of an  arbitrarily chosen radio candidate counterpart within the INTEGRAL large error circle. However, we note that this interpretation is inconsistent with our reported properties on IGR~J21268+6203. Specifically, the short hard X-ray outburst duration and the lack of persistent  hard X-ray emission are incompatible with the typical behavior of non blazar-like AGNs,  which are known to be persistent hard X-ray emitters when detected by INTEGRAL. Moreover the candidate radio counterpart, previously suggested as  candidate AGN, was not detected in  X-rays despite being observed for the first time in this band, casting serious doubt on its association with IGR J21268+6203. In this context, we note that  only a small fraction of AGN are intrinsically X-ray weak. Systematic studies from large surveys show that only $\sim$5\% are underluminous  in the X-rays by a factor of  $>$6 compared to the X-ray level expected from their ultraviolet emission \citep{xingting2020}. Such extreme X-ray faintness is therefore rare among the class of AGN. Thus, the absence of an X-ray detection for the previously proposed AGN counterpart of IGR J21268+6203 is statistically unlikely for a typical AGN, and  favors a Galactic origin for the true counterpart.

Given its proximity and modest height above the plane, IGR J21268+6203 is plausibly a nearby
flaring star. Indeed the optical/infrared  counterpart is a late-type K0 star. The soft and thermal quiescent X-ray spectrum measured with \sw/XRT and the corresponding low  X-ray 
luminosity of $\sim$4.8$\times$10$^{29}$ erg s$^{-1}$ both fully agrees  with the X-ray 
characteristics  of flaring late-type stars. 
Also the very low ratio of X-ray to optical flux \citep{Maccacaro1988} confirms that the X-ray emission comes from the K-type star (log$(\frac{f_X}{f_{\mathrm{opt}}})=-4.9$, using V=7.6 mag  and a  flux of  5$\times$10$^{-14}$ erg cm$^{-2}$ s$^{-1}$ (0.3-3.5 keV) measured with \sw/XRT).
As for the INTEGRAL detection in outburst, both the  duration and average X-ray luminosity are consistent with a flaring star nature \citep{zhao2024}
and match expectations for stellar flares detected by INTEGRAL in the hard X-ray 
band \citep{sgu2016ate}.


\subsection{IGR J02447+7046}

IGR J02447+7046 is a hard X-ray transient detected by INTEGRAL in outburst only once, in November 2003, over the 15-year IBIS/ISGRI archive. The outburst duration is constrained to the range 5–30 days, as only the early phase of the activity was covered by INTEGRAL before the source moved out of its FoV. The average 18-60 keV flux is $\sim$1.0$\times$10$^{-10}$ erg cm$^{-2}$ s$^{-1}$, corresponding to a luminosity of $\sim$5.8$\times$10$^{34}$ erg s$^{-1}$ at a distance of 2.2 kpc, although the source likely reached a higher flux during the unobserved interval. Unfortunately, no soft or hard X-ray all-sky monitors (such as MAXI or \sw/BAT) were operational at the time to constrain a possible brighter phase. Despite the extensive INTEGRAL monitoring (effective exposure $\sim$1.1 Ms), no persistent hard X-ray emission was detected. We derive a 3$\sigma$  upper limit in the 20–40 keV band of 0.4 mCrab or 3$\times$10$^{-12}$ erg cm$^{-2}$ s$^{-1}$ (1.7$\times$10$^{33}$ erg s$^{-1}$ at 2.2 kpc). 

To search for the most likely X-ray counterpart, we analyzed  \sw/XRT observations covering the entire INTEGRAL error circle. 
Only one \sw/XRT source is detected within it, hence we propose it
as the candidate soft X-ray counterpart.
Within its arcsecond-sized  error circle, we identified a single optical/infrared counterpart located at $\sim$2.2 kpc. Its photometric properties are consistent with  an early-type star. Remarkably, this low-energy counterpart is also detected as a very bright radio source at 20, 92, and 200 cm.

For completeness, we note that a Seyfert AGN had previously been proposed in the literature as the counterpart of IGR J02447+7046. The AGN is detected in all 
\sw/RT observations, however its position strikingly lies outside the INTEGRAL error circle at 95\% confidence. Moreover, its persistent X-ray nature is at odds with the transient hard X-ray behavior as well as the stringent INTEGRAL upper limit on the persistent emission of IGR J02447+7046. In addition, its redshift (z=0.306) is significantly  higher than typical values for Seyfert AGNs firmly detected by INTEGRAL (average z$\sim$0.033), it would be  difficult to justify  
IGR J02447+7046 as outlier. All these findings suggest that the Seyfert AGN is too distant to be detected at INTEGRAL hard X-ray energies (see details in section 4.1 for the similar cae of IGR~J16426+6536). Indeed, IGR J02447+7046 is not listed in the 
\sw/BAT 157-months survey hard X-ray catalog, which includes nearly 1,000 Seyfert AGNs. Taken at face value, our results rule out the AGN scenario for IGR J02447+7046.

We propose that IGR J02447+7046 is a Galactic massive binary system with a neutron star, based on our collected results. Considering that the source distance is  $\sim$2.2 kpc and it is  located at a Galactic latitude of $\sim$10$^{\circ}$, this places it about 382 pc above the Galactic plane. Such a vertical height remains compatible with a Galactic origin, especially for high-mass X-ray binaries (HMXBs) that may have been displaced from the plane by supernova kicks. The transient hard X-ray emission of IGR J02447+7046, lasting between 5 and 30 days and having a lower limit luminosity in outburst of $\sim$5.8$\times$10$^{34}$ erg s$^{-1}$, is in agreement  with the typical outburst characteristics of Be HMXBs \citep{reig2011},  as opposed to the much shorter activity typically observed in Supergiant Fast X-ray Transients \citep{sgu2006}. We pinpointed an early-type optical/infrared counterpart which 
supports this classification. Although there is no evidence for a soft X-ray variability during the very few snapshot taken by \sw/XRT at the source location, its hard spectrum in the 0.3-10 keV energy range is compatible with a massive X-ray binary.

Notably, the association of the proposed soft X-ray and optical counterpart with a very bright radio source is intriguing, and leads us to propose a specific type of massive binary, i.e. a $\gamma$-ray binary composed of a young pulsar orbiting a Be star. By adopting the three available radio flux measurements  at 20, 92, and 200 cm, we calculate a power law spectral slope $\alpha$=$-$0.56 (where the flux density is S$_\nu$$\propto\nu^{\alpha}$). This value is  typical of optically thin  synchrotron radio emission and is consistent with a non-thermal origin,  such as a pulsar wind shock  in a $\gamma$-ray binary.
An alternative explanation involving radio emission from a relativistic jet is unlikely, since the radio spectrum in this scenario is usually flatter (with $\alpha$ $\sim$ 0.0). 
Moreover, if we extrapolate the power law radio spectrum ($\alpha$=$-$0.56) to 6 GHz, we obtain a flux density of $\approx$80~mJy, translating into a radio luminosity of L$_{R}$$\sim$3$\times10^{30}$~erg~s$^{-1}$ (at 2.2 kpc).  
This 6 GHz luminosity would  be comparable to jet emission only in the case of  black hole X-ray binaries with  X-ray luminosities exceeding 10$^{36}$~erg s$^{-1}$ (0.5-10 keV),  assuming the observed correlation for such systems in the radio/X-ray plane \citep{vdEijnden2021}. Moreover, X-ray transient black hole binaries host low mass companions, at odds with the early-type optical counterpart.

On the other hand, $\gamma$-ray binaries composed by a young pulsar orbiting a Be star typically show high radio luminosities and relatively low X-ray luminosities, characteristics that are more consistent with our proposed counterpart of IGR~J02447+7046. A notable example is  PSR~B1259-63, which  hosts a 47.76~ms pulsar orbiting the Be star LS~2883 in a 3.4 yrs highly eccentric orbit  (e=0.87). Located at a similar distance (2.4 kpc), this system shows variable emission across the electromagnetic spectrum, from radio to TeV energies, modulated by the orbital period. It also displays a flux density of a few tenths mJy  at frequencies $\sim$6 GHz, a similar radio spectral index, a hard X-ray spectrum below 10 keV (power law photon index of 1.5-1.9) and low X-ray luminosities in the range of $\sim$10$^{33}$-10$^{34}$~erg~s$^{-1}$ in soft X-rays; \citealt{Kaspi1995, Chernyakova2024}). 
Since the radio emission is produced in the shock between the pulsar wind and the Be star's outflow along the orbit, it can be highly  variable depending from the specific properties of the system and from its orbital geometry. 
We note that no known MeV-GeV sources listed in published Fermi/LAT catalogs are positionally consistent with  NVSS~J024443+704946. As for the TeV band, to date the sky region of IGR J02447+7046 has never been covered by MAGIC or VERITAS observations, while the source is not visible to H.E.S.S. due to its high northern declination. 
Therefore, a more detailed and  close comparison with known $\gamma$-ray binaries might be premature
at this stage.  To better understand the nature and origin of the radio emission detected from 
IGR J02447+7046, dedicated follow-up observations in the millimeter and radio bands are 
encouraged.

\subsection{Additional remarks on the INTEGRAL sources}

\begin{figure}
\begin{center}
\includegraphics[height=6.8 cm, angle=0]{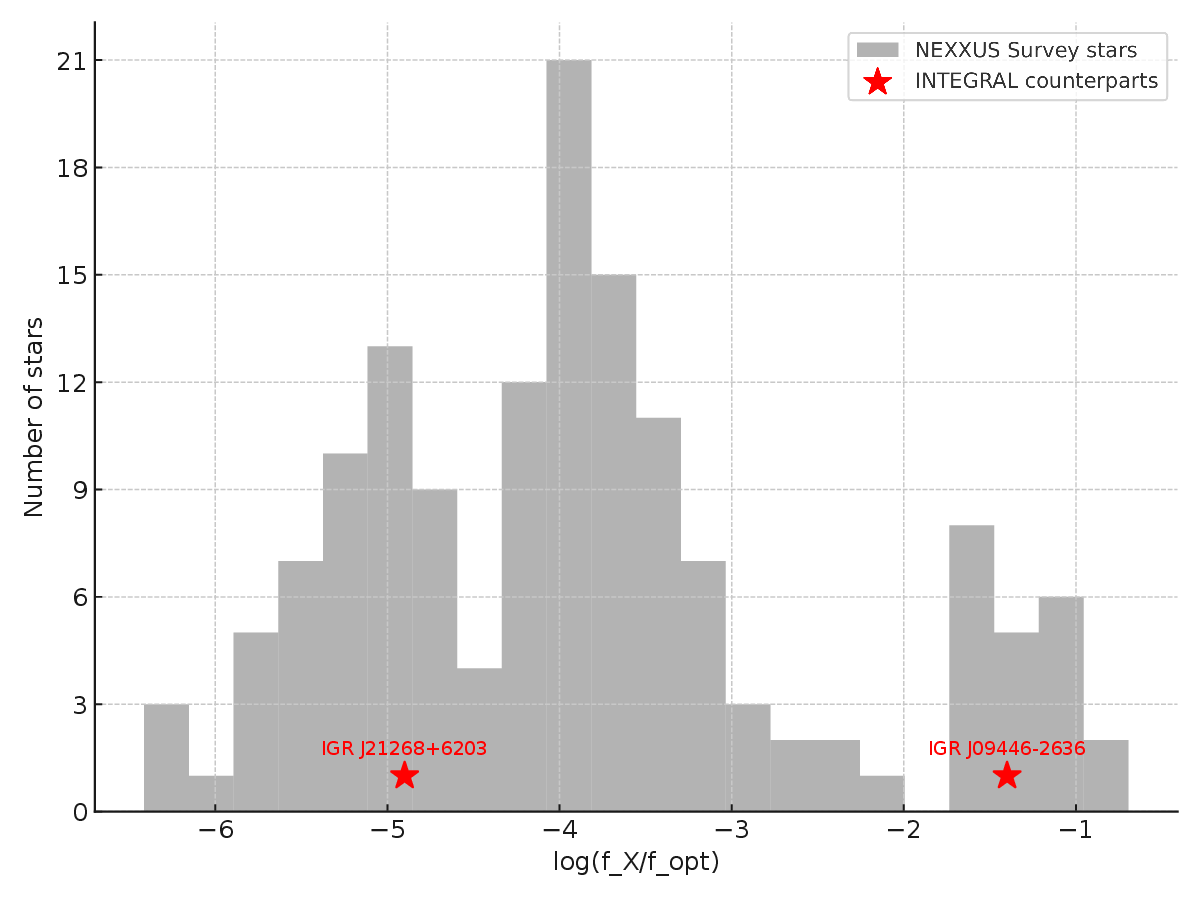}
\caption{Distribution of $\log(f_{\rm X}/f_{\rm opt})$ for nearby late-type stars 
from the NEXXUS Survey sample \citep{Schmitt2004}, computed following 
\citet{Maccacaro1988}. The proposed X-ray counterparts of 
IGR~J09446$-$2636  and IGR~J21268+6203 are shown as red stars. Both sources 
lie within the stellar locus.}
\end{center}
\end{figure}

The robustness of the four INTEGRAL transient sources reported in the IBIS catalog by  Bird et al. (2016) can be assessed as follow.  As stated  in the catalog, simulations based on randomized and inverted light curves show that, for transient sources detected with the bursticity method, the mean false positive rate alarm probability per light curve above 6$\sigma$ significance lies in the range  1\% (statistical best case) and 3.5\% (systematic worst case). When requiring temporal coincidence in more than one energy band (the criterion adopted in the catalog),  the statistical probability falls to $<$1\% even above 4.5$\sigma$. The final overall false positive rate for sources detected with the bursticity method is pessimistically assumed to be $\sim$1\% after accounting for possible systematic effects,  and is further constrained by manual image/light-curve inspection to reject ghosts and noisy intervals. In particular, the false positive probability for transient sources detected with the bursticity at lower significance ($<$6$\sigma$) and on shorter timescale is higher than the cited overall final level. Warning flags are appended to such sources in the table catalog to indicate detections subject to higher false positive rate. We note that, among the four INTEGRAL transients studied in this work, IGR~J21268+6203 is the only one indicated with a warn flag in the IBIS catalog. However, we point out that its significance detection (5.9$\sigma$) is very slightly below the mentioned 6$\sigma$ threshold.  All the above assumptions support the conclusion that at least three of the four INTEGRAL transients are genuine astrophysical sources, with a very strong likelihood that all four are real.

To strengthen the rejection of the AGNs as candidate counterparts, we investigated their long-term behaviour.  For the case of  IGR J16426+6536,  IGR J02447+7046 and IGR J09446$-$2636, the associated AGN is characterized by persistent soft X-ray emission (expected for Seyfert galaxies) with no sign of X-ray flaring behaviour, while the INTEGRAL detections correspond to isolated, short hard X-ray outbursts with no persistent hard X-ray emission. We note that for IGR~J21268+6203 the associated radio candidate AGN is undetected in X-rays.  Overall, none of the four AGNs are reported in the \sw\/BAT 157-month catalog, which contains $\sim$1,000 Seyferts \citep{lien2025}. This behaviour contrasts with that of known $\gamma$-ray--flaring AGN, including narrow-line Seyfert 1 galaxies, where high-energy outbursts are always accompanied by strong, multi-wavelength variability and persistent X-ray detection.  Moreover, $\gamma$-ray emitting NLS1s are rare, and even among them only a few display extreme high energy  flares \citep[e.g.][]{dammando2015}. Consequently, the available long-term constraints coupled with the absence of multi-band flaring episodes disfavor an AGN association for the four INTEGRAL sources.

To place the two proposed X-ray counterparts of IGR~J09446$-$2636 and IGR~J21268+6203
in the broader context of stellar X-ray emitters, we compared them with the NEXXUS survey
\citep{Schmitt2004}, which is based on ROSAT observations and provides a nearly complete census of nearby late-type stars. NEXXUS characterizes the coronal X-ray properties of these stars, making it an ideal reference sample. The distribution of  $\log(f_{\rm X}/f_{\rm opt})$ derived from NEXXUS shows that the proposed INTEGRAL counterparts of IGR~J09446$-$2636 and 
IGR~J21268+6203 fall well within the stellar locus (see Fig. 6). This demonstrates that their X-ray to optical flux ratios are fully consistent with  stellar coronal emission, reinforcing their classification. 

To further assess the robustness of our proposed X-ray counterparts, we estimated the probability of a chance coincidence for IGR~J16426+6536 and IGR~J09446$-$2636. We excluded from the calculation the remaining cases of IGR~J02447+7046 and IGR~J21268+6203 because inside their  ISGRI error circle  there is only a single \sw/XRT  counterpart, since the bright X-ray AGN lies outside of it in the case of  IGR~J02447+7046 while the candidate radio AGN is not detected in X-ray in the case of IGR~J21268+6203.
The probability that the nearest unrelated X-ray source lies within angular distance r 
is $P$=1-e$^{-\pi\rho r^2}$.
We adopted a value of the surface density of background X-ray sources ($\rho$) equal to $\rho$=8 deg$^{-2}$ and $\rho$=30 deg$^{-2}$ for IGR~J09446$-$2636 and 
IGR~J16426+6536, respectively \citep{mateos2008}. 
We used the observed source flux in the energy band 0.5-2 keV  (7.1$\times$10$^{-14}$ erg cm$^{-2}$ s$^{-1}$) for  IGR~J09446$-$2636 and in the 2-10 keV energy range (4.7$\times$10$^{-14}$ erg cm$^{-2}$ s$^{-1}$) for IGR~J16426+6536. We obtained chance association probabilities of $\sim$28\% for IGR~J16426+6536 (r=3$\arcmin$.5) and $\sim$19\% for IGR~J09446$-$2636 (r=5$\arcmin$.47).  
At face value, such probabilities  appear non-negligible and could suggest, if regarded on their own, a spurious association. However, when combined with the many independent evidences supporting the associations, and with the spectral and temporal arguments that disfavor the brighter AGN as possible counterparts,  our proposed X-ray counterparts are  plausible.


\section{Conclusions}
We have carried out a comprehensive analysis of four  INTEGRAL hard X-ray sources, whose X-ray characteristics were previously largely unknown. These sources had been tentatively classified as AGN in earlier works, but such associations were based on unconfirmed evidence. In this work we propose alternative soft X-ray counterparts, pointing to a Galactic nature instead.

By combining soft and hard X-ray observations, we have been able to characterize their X-ray characteristics in greater detail and to pinpoint the most likely X-ray counterpart. Through a multiwavelength approach, we identified  their optical and infrared counterparts. 
All the collected results and the derived properties of these counterparts 
point to a Galactic origin for all four INTEGRAL hard X-ray sources. Specifically, we find that one is  likely a VFXT or a classical LMXB (IGR J16426+6536), one is best interpreted as a massive binary with a Be companion, in particular a $\gamma$-ray binary (IGR~J02447+7046) and finally two are consistent with a flaring star nature (IGR J09446$-$2636 and IGR J21268+6203).


\begin{acknowledgements}
We thank the  referee for the prompt and constructive comments. 
This work is based on observations performed with \xmm, \swift\ and \int\  satellites. 
We made use of the High Energy Astrophysics Science Archive Research Center (HEASARC), a service of the Astrophysics Science Division at NASA/GSFC. This research has made use of data obtained from the 4XMM XMM-Newton serendipitous source catalogue compiled by the XMM-Newton Survey Science Centre consortium. 
This work has made use of data from the European Space Agency (ESA) mission Gaia (https://www.cosmos.esa.int/gaia), processed by the Gaia Data Processing and Analysis Consortium (DPAC, https://www.cosmos.esa.int/web/gaia/dpac/consortium). This work made use of data supplied by the UK Swift Science Data Centre at the University of Leicester (\cite{Goad2007}, \cite{evans2009}).We acknowledge funding from the grant entitled ``Bando Ricerca Fondamentale INAF 2023". 
\end{acknowledgements}


\bibliographystyle{aa}
\bibliography{biblio_ls}



%

\clearpage

\begin{appendix}

\section{Additional figures}

\begin{figure} 
\begin{center}
\includegraphics[height=7 cm, angle=-90]{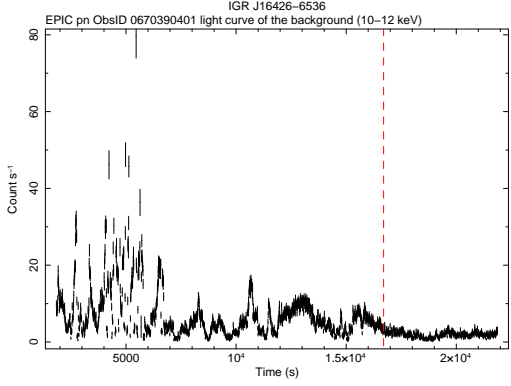}
\caption{\xmm\ EPIC pn background light curve (10-12 keV) IGR J16426+6536 extracted from a large source-free region of ObsID~0670390401 (bin time of 10 s). For any subsequent data analysis and extraction of products, we selected only the last $\sim$5~ks, right of the dashed, red vertical line.} 
\label{fig:xmmbgflares}
\end{center}
\end{figure} 

\begin{figure}
\begin{center}
\label{fig:src2_pow_igr16426}
\includegraphics[height=7 cm, angle=-90]{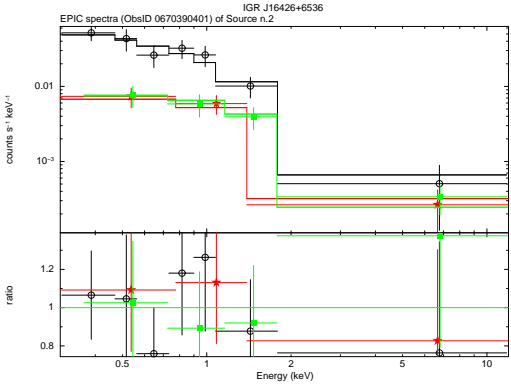}
\caption{\xmm\ EPIC spectra of X-ray source n.2 (ObsID~0670390401) counterpart of IGR J16426+6536. The top panel shows the data fitted with an absorbed power law model, while the bottom panel the ratio of the data to the model. Open (black) circles mark the pn spectrum, solid (red) stars indicates the MOS~1 spectrum and solid (green) squares mark the MOS~2 data points. The spectra have been severely binned for graphical purposes only.
} 
\end{center}
\end{figure} 

\begin{figure} 
\begin{center}
\includegraphics[height=7 cm, angle=-90]{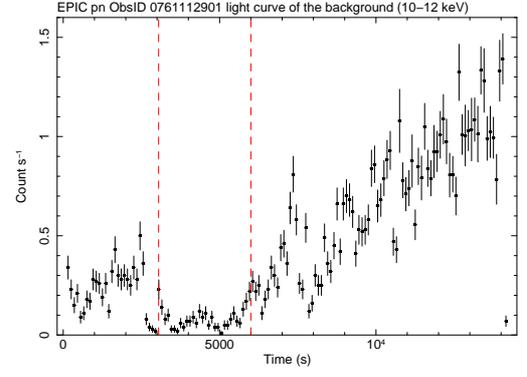}
\caption{\xmm\ EPIC pn background light curve (10-12 keV)  of IGR J09446-2636  extracted from a large source-free region of ObsID~0761112901. For any subsequent data analysis and extraction of products, we selected the time interval with a duration of $\sim$3~ks within the two red, dashed vertical lines (time bin = 100 s).} 
\label{fig:xmmbgflares-igr09446}
\end{center}
\end{figure} 

\begin{figure}
\begin{center}
\label{fig:src2_apec_igr09446_pnm12}
\includegraphics[height=7cm, angle=-90]{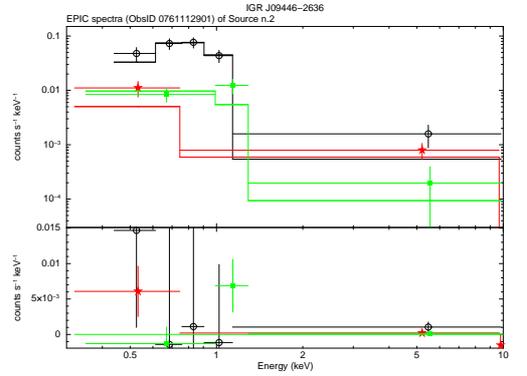}
\caption{\xmm\ EPIC spectra of X-ray source n.2 (ObsID~0761112901) counterpart of IGR J09446-2636. The top panel shows the data fitted with an absorbed thermal plasma model, while the bottom panel the residuals. Open (black) circles mark the pn spectrum, solid (red) stars indicates the MOS~1 spectrum and solid (green) squares mark the MOS~2 data points. The spectra have been severely binned for graphical purposes only.} 
\end{center}
\end{figure} 

\begin{figure}
\begin{center}
\label{fig:sw_spec_igr21268}
\includegraphics[height=7 cm, angle=-90]{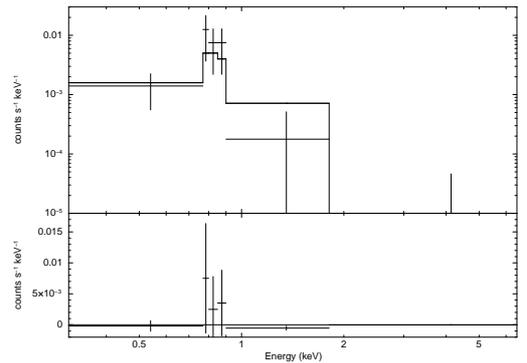}
\caption{\sw/XRT spectrum of the newly proposed counterpart of IGR~J21268+6203 from ObsID 45405006, as reported in  Table~\ref{tab:xmmxrtspec}.  The residuals of the data with respect to the model are shown in the bottom panel. The spectrum has been severely binned for graphical purposes only.}

\end{center}
\end{figure}

\end{appendix}

\end{document}